# The geometric algebra lift of qubits via basic solutions of Maxwell equation

Alexander SOIGUINE[1]

[1] SOiGUINE Quantum Computing, Aliso Viejo, CA 92656, USA

http://soiguine.com

Email address: alex@soiguine.com

**Abstract:** Conventional quantum mechanical qubits can be lifted to states as $G_3^+$ valued operators that act on observables [1]. That operators may be implemented via the two types of Maxwell equations' solution polarizations [2]. Solution of Maxwell equation in geometric algebra formalism gives g-qubits which are exact lifts of conventional qubits. Therefore, it unambiguously reveals actual meaning of complex parameters of qubits of the commonly accepted Hilbert space quantum mechanics and, particularly, directly demonstrates the option of instant nonlocality of states.

## 1. Introduction

As I straightly declared, valuable mathematical models applicable to quantum processes can only appear when formalized in consistent, adequate terms. Inability to realize that 'complex' numbers should be real geometrical objects of the three-dimensional world resulted in more than a century bewitchment about non-observable observables in commonly used mathematical structure of quantum mechanics. And, wave-particle mysterious dualism follows mainly from the lack of clear logical distinguishing between operators and operands [3].

Recent results on solution of Maxwell equation(s) [2]

$$(\partial_t + \nabla)F = 0 \qquad (1.1)$$

in geometric algebra terms showed that the general solution is linear combination of two basic solutions:

- $F_+ = (\vec{E}_0 + I_3 \vec{H}_0) exp[I_S(\omega t - \vec{k}_+ \cdot \vec{r})]$ , where the vector $\hat{k}_+ = I_3 I_S$[1] is orthogonal to the given plane $S$ and the triple of vectors $\{\hat{E}, \hat{H}, \hat{k}_+\}$ is right hand screw oriented, that's rotation of $\hat{E}$ to $\hat{H}$ by angle less than $\pi$ moves the *right hand screw* in the direction of $\vec{k}_+ = |\vec{k}| I_3 I_S$.

---

[1] For any vector the notation is used: $\hat{a} = \vec{a}/|\vec{a}|$

- $F_- = (\vec{E}_0 + I_3\vec{H}_0)exp[I_S(\omega t - \vec{k}_- \cdot \vec{r})]$, where $\hat{k}_- = -I_3 I_S$ is orthogonal to the given plane $S$ but the triple of vectors $\{\hat{E}, \hat{H}, \hat{k}_-\}$ is left hand screw oriented, that's rotation of $\hat{E}$ to $\hat{H}$ by angle less than $\pi$ moves the *left hand screw* in the direction of $\vec{k}_- = -|\vec{k}|I_3 I_S$ or, equivalently, moves the *right hand screw* in the opposite direction, $-\vec{k}_-$.
- $\vec{E}_0$ and $\vec{H}_0$, initial values of $\vec{E}$ and $\vec{H}$, are arbitrary mutually orthogonal vectors, of the same length and lying on the plane $S$:
$$F_0 = \vec{E} + I_3\vec{H}\Big|_{t=0,\vec{r}=0} = \vec{E}\Big|_{t=0,\vec{r}=0} + I_3\vec{H}\Big|_{t=0,\vec{r}=0} = \vec{E}_0 + I_3\vec{H}_0$$
- The lengths $|\vec{k}_\pm|$ of both wave vectors are the same and equal to angular frequency $\omega$.
- $I_S$ is unit bivector in plane $S$ and $I_3$ is unit value right-hand oriented volume in three-dimensional Euclidean space.

The above two circular polarized electromagnetic waves are the only two types of waves available as solutions of Maxwell equations in free space in geometric algebra terms.

Equation (1.1) is a linear one, thus for arbitrary scalars[2] $\lambda$ and $\mu$:

$$\lambda F_+ + \mu F_- = (\vec{E}_0 + I_3\vec{H}_0)e^{I_S\omega t}\left(\lambda e^{-I_S\omega[(I_3 I_S)\cdot\vec{r}]} + \mu e^{I_S\omega[(I_3 I_S)\cdot\vec{r}]}\right) \quad (1.2)$$

is also solution. Since in terms of the current geometrical algebra approach the items of the type $e^{I_S\varphi}$ are states, named g-qubits [1], then, particularly, the sum in the second parenthesis is weighted linear combination of two states with the same, by absolute value, phase, but with opposite sense of the plane $S$ orientation. The two states are strictly coupled, entangled if you prefer, since the bivector plane should be the same, up to orientation, in both solutions. That means that entanglement is inherent to the two coupled solutions of the Maxwell equation when the latter is formulated in geometric algebra terms. Circular polarizations received as solutions of Maxwell equation (1.1) is an excellent choice to have g-qubits in a lab.

## 2. Conventional qubits and the two Maxwell equation basic solutions

Qubit in conventional quantum mechanics is an element of two-dimensional Hilbert space of the form:

$$c_1|0\rangle + c_2|1\rangle = c_1\begin{pmatrix}1\\0\end{pmatrix} + c_2\begin{pmatrix}0\\1\end{pmatrix}$$

where $c_1$ and $c_2$ are complex numbers in their usual meaning. When complex numbers are generalized to explicit geometrical three dimensional objects, elements of $G_3^+$, even subalgebra of geometric algebra $G_3$, [4], [5], then a qubit, for some chosen triple of basis bivectors in three dimensions $\{B_1, B_2, B_3\}$, gets lift to $G_3^+$, defined up to arbitrary permutation of elements $\{B_1, B_2, B_3\}$, for example:

$$c_1|0\rangle + c_2|1\rangle \equiv (\alpha + i\beta_1)|0\rangle + (\beta_3 + i\beta_2)|1\rangle \Longrightarrow \alpha + \beta_1 B_1 + \beta_2 B_2 + \beta_3 B_3$$

---

[2] I should remind that in the current approach scalars are real numbers, complex valued scalars do not make sense

It follows that qubit state $c_1|0\rangle = (c_1^1 + ic_1^2)|0\rangle$ has the lift $c_1^1 + c_1^2 B_1$ and qubit state $c_2|1\rangle = (c_2^1 + ic_2^2)|1\rangle$ has the lift $c_2^2 B_2 + c_2^1 B_3 = (c_2^1 + c_2^2 B_1)B_3$ since basis bivectors satisfy multiplication rules (in the righthand screw orientation of $I_3$):

$$B_1 B_2 = -B_3,\ B_1 B_3 = B_2,\ B_2 B_3 = -B_1 \quad (2.1)$$

Measurement of an observable with the bivector component $C_1 B_1 + C_2 B_2 + C_3 B_3$ by the state $c_1^1 + c_1^2 B_1$ does not change the observable component $C_1 B_1$, [1], [5], while the state $(c_2^1 + c_2^2 B_1)B_3$ additionally flips the component around $B_1$ resulting in $-C_1 B_1$. That is the actual meaning of the Hilbert space states $|0\rangle$ and $|1\rangle$ when lifted to $G_3^+$.

Let's analyze two items belonging to $F_+$ and $F_-$: $\vec{E}_0 e^{I_S \omega(t \pm [(I_3 I_S) \cdot \vec{r}])}$ and $I_3 \vec{H}_0 e^{I_S \omega(t \pm [(I_3 I_S) \cdot \vec{r}])}$. Element $I_3 \vec{H}_0$ is bivector. Denote it by $I_3 \vec{H}_0 \equiv B_0$. Thus, we have $\vec{E}_0 e^{I_S \omega(t \pm [(I_3 I_S) \cdot \vec{r}])}$ and $B_0 e^{I_S \omega(t \pm [(I_3 I_S) \cdot \vec{r}])}$. The latter has, up to the scalar value of bivector $B_0$, the form of a state from $G_3^+$. Bivector $B_0$ is orthogonal to bivector $I_S$ since $\vec{H}_0$ lies in the plane $S$. That means that $I_{B_0} e^{I_S \omega(t \pm [(I_3 I_S) \cdot \vec{r}])}$ is a state corresponding to some $c|1\rangle$ from conventional quantum mechanics. In a measurement of any observable from $G_3$ it flips, after rotation in $S$, the result around the plane $S$. It follows from the fact that if we take a triple of unit value bivectors $\{I_S, I_{B_0}, I_{E_0}\}$ then, due to the mutual properties of $\hat{E}, \hat{H}, \hat{k}_+$ described in Introduction, the triple can be considered as bivector basis with multiplication rules:

$$I_S I_{B_0} = -I_{E_0},\ I_S I_{E_0} = I_{B_0},\ I_{B_0} I_{E_0} = -I_S$$

Then, for example:

$$I_{B_0} e^{I_S \omega(t \pm [(I_3 I_S) \cdot \vec{r}])} \equiv I_{B_0} e^{I_S \varphi} = I_{B_0}(\cos\varphi + I_S \sin\varphi) = I_{B_0} \cos\varphi + I_{E_0} \sin\varphi$$

and measurement of observable $I_S$ is:

$$(-I_{B_0} \cos\varphi - I_{E_0} \sin\varphi) I_S (I_{B_0} \cos\varphi + I_{E_0} \sin\varphi) = (-I_{E_0} \cos\varphi + I_{B_0} \sin\varphi)(I_{B_0} \cos\varphi + I_{E_0} \sin\varphi)$$
$$= -I_S \cos^2\varphi - \sin\varphi \cos\varphi + \sin\varphi \cos\varphi - I_S \sin^2\varphi = -I_S$$

If we take an observable as arbitrary bivector expanded in basis $\{I_S, I_{B_0}, I_{E_0}\}$, say $C_1 I_S + C_2 I_{B_0} + C_3 I_{E_0}$, then its measurement by the state $I_{B_0} e^{I_S \omega(t \pm [(I_3 I_S) \cdot \vec{r}])}$ gives [5]:

$$-C_1 I_S + (C_2 \cos 2\varphi + C_3 \sin 2\varphi) I_{B_0} + (C_2 \sin 2\varphi - C_3 \cos 2\varphi) I_{E_0}$$

with $\cos 2\varphi = \cos 2\omega(t \pm [(I_3 I_S) \cdot \vec{r}])$ and $\sin 2\varphi = \sin 2\omega(t \pm [(I_3 I_S) \cdot \vec{r}])$. This is rotation around the axis orthogonal to $S$ by the angle $2\omega(t \pm [(I_3 I_S) \cdot \vec{r}])$ with subsequent flip of the result over $I_S$ that particularly changes the sense of the $I_S$ component.

The $\vec{E}_0 e^{I_S \omega(t \pm [(I_3 I_S) \cdot \vec{r}])}$ element is different. Action of vector $\vec{E}_0$ on observable $I_S$ is reflection of $I_S$ relative to the plane orthogonal to that vector [5]. The vector is orthogonal to $I_S$ so the reflection is just $I_S$ itself. Subsequent action of $e^{I_S \omega(t \pm [(I_3 I_S) \cdot \vec{r}])}$ does not change $I_S$ because it is rotation of $I_S$ in its own plane. Similar to the calculations for $I_{B_0} e^{I_S \omega(t \pm [(I_3 I_S) \cdot \vec{r}])}$, the action of $e^{I_S \omega(t \pm [(I_3 I_S) \cdot \vec{r}])}$ on $C_1 I_S + C_2 I_{B_0} + C_3 I_{E_0}$ gives [5]:

$$C_1 I_S + (C_2 \cos 2\varphi - C_3 \sin 2\varphi) I_{B_0} + (C_2 \sin 2\varphi + C_3 \cos 2\varphi) I_{E_0}$$

The component $C_1 I_S$ is not changed in the measurement and two other bivector components $C_2 I_{B_0}$ and $C_3 I_{E_0}$, orthogonal to $I_S$ and to each other, make usual rotation around the axis orthogonal to $S$ by angle $2\omega(t \pm [(I_3 I_S) \cdot \vec{r}])$.

If we separate, as in (1.2), two partial solutions with opposite orientations of $I_S$ in the space dependent exponents, then it follows the next statement:

General solution of Maxwell equation is, up to constant value amplitudes, arbitrary linear combination of two geometric algebra $G_3^+$ elements, corresponding to conventional quantum mechanics states of the types $c_1|0\rangle$ and $c_2|1\rangle$:

$$e^{I_S \omega(t - [(I_3 I_S) \cdot \vec{r}])} + I_{B_0} e^{I_S \omega(t - [(I_3 I_S) \cdot \vec{r}])}$$
$$= \cos\omega(t - [(I_3 I_S) \cdot \vec{r}]) + I_S \sin\omega(t - [(I_3 I_S) \cdot \vec{r}]) + I_{B_0} \cos\omega(t - [(I_3 I_S) \cdot \vec{r}])$$
$$+ I_{E_0} \sin\omega(t - [(I_3 I_S) \cdot \vec{r}])$$

$$e^{I_S \omega(t + [(I_3 I_S) \cdot \vec{r}])} + I_{B_0} e^{I_S \omega(t + [(I_3 I_S) \cdot \vec{r}])}$$
$$= \cos\omega(t + [(I_3 I_S) \cdot \vec{r}]) + I_S \sin\omega(t + [(I_3 I_S) \cdot \vec{r}]) + I_{B_0} \cos\omega(t + [(I_3 I_S) \cdot \vec{r}])$$
$$+ I_{E_0} \sin\omega(t + [(I_3 I_S) \cdot \vec{r}])$$

Both can be, with included normalization factor $\frac{1}{\sqrt{2}}$, written as geometric algebra states of $G_3^+$ in the usual form $\alpha + \beta I_{Plane}$, where $\alpha^2 + \beta^2 = 1$ and $I_{Plane}$ – some unit vale bivector in the three dimensions:

$$\frac{1}{\sqrt{2}} \cos\omega(t - [(I_3 I_S) \cdot \vec{r}])$$
$$+ \frac{\sqrt{1 + \sin^2\omega(t - [(I_3 I_S) \cdot \vec{r}])}}{\sqrt{2}} \left[ I_S \frac{\sin\omega(t - [(I_3 I_S) \cdot \vec{r}])}{\sqrt{1 + \sin^2\omega(t - [(I_3 I_S) \cdot \vec{r}])}} + I_{B_0} \frac{\cos\omega(t - [(I_3 I_S) \cdot \vec{r}])}{\sqrt{1 + \sin^2\omega(t - [(I_3 I_S) \cdot \vec{r}])}} \right.$$
$$+ I_{E_0} \left. \frac{\sin\omega(t - [(I_3 I_S) \cdot \vec{r}])}{\sqrt{1 + \sin^2\omega(t - [(I_3 I_S) \cdot \vec{r}])}} \right] \quad (2.2)$$

$$\frac{1}{\sqrt{2}} \cos\omega(t + [(I_3 I_S) \cdot \vec{r}])$$
$$+ \frac{\sqrt{1 + \sin^2\omega(t + [(I_3 I_S) \cdot \vec{r}])}}{\sqrt{2}} \left[ I_S \frac{\sin\omega(t + [(I_3 I_S) \cdot \vec{r}])}{\sqrt{1 + \sin^2\omega(t + [(I_3 I_S) \cdot \vec{r}])}} + I_{B_0} \frac{\cos\omega(t + [(I_3 I_S) \cdot \vec{r}])}{\sqrt{1 + \sin^2\omega(t + [(I_3 I_S) \cdot \vec{r}])}} \right.$$
$$+ I_{E_0} \left. \frac{\sin\omega(t + [(I_3 I_S) \cdot \vec{r}])}{\sqrt{1 + \sin^2\omega(t + [(I_3 I_S) \cdot \vec{r}])}} \right] \quad (2.3)$$

since $1 - \frac{1}{2}\cos^2\omega(t \pm [(I_3 I_S) \cdot \vec{r}]) = \frac{1}{2}\left(1 + \sin^2\omega(t \pm [(I_3 I_S) \cdot \vec{r}])\right)$

As any g-qubit $\alpha + \beta_1 B_1 + \beta_2 B_2 + \beta_3 B_3$ representable as linear combination, composition, of two operators, the lifts of a conventional qubit components $c_1|0\rangle$ and $c_2|1\rangle$

$$\alpha + \beta_1 B_1 + \beta_2 B_2 + \beta_3 B_3 = \alpha + \beta_1 B_1 + (\beta_3 + \beta_2 B_1) B_3$$

where $\alpha + \beta_1 B_1$ executes rotation of an observable in the plane of $B_1$ by angle $\cos^{-1}\alpha$ and $(\beta_3 + \beta_2 B_1) B_3$ executes rotation in the same plane by angle $\cos^{-1}\beta_3$ with subsequent flip of the result over the plane $B_1$, the above two $G_3^+$ state operators can be expressed in the same way:

$$\frac{1}{\sqrt{2}}\cos\omega(t-[(I_3I_S)\cdot\vec{r}]) + \frac{1}{\sqrt{2}}\sin\omega(t-[(I_3I_S)\cdot\vec{r}])I_S$$
$$+\frac{1}{\sqrt{2}}(\sin\omega(t-[(I_3I_S)\cdot\vec{r}]) + \cos\omega(t-[(I_3I_S)\cdot\vec{r}])I_S)I_{E_0}$$

$$\frac{1}{\sqrt{2}}\cos\omega(t+[(I_3I_S)\cdot\vec{r}]) + \frac{1}{\sqrt{2}}\sin\omega(t+[(I_3I_S)\cdot\vec{r}])I_S$$
$$+\frac{1}{\sqrt{2}}(\sin\omega(t+[(I_3I_S)\cdot\vec{r}]) + \cos\omega(t+[(I_3I_S)\cdot\vec{r}])I_S)I_{E_0}$$

## 3. Transformation of the Maxwell's equation g-qubit basic solutions via Hamiltonian

Suppose we have conventional Hamiltonian of general form that can act on qubits:

$$\begin{pmatrix} a+b & c-id \\ c+id & a-b \end{pmatrix} \quad (3.1)$$

Its geometric algebra lift is (see, for example, [5], p.34):

$$\begin{pmatrix} a+b & c-id \\ c+id & a-b \end{pmatrix} \Rightarrow a + I_3(bB_1 + cB_2 + dB_3) = a + I_3B_H \equiv H$$

with an arbitrary taken bivector basis $\{B_1, B_2, B_3\}$ in three dimensions satisfying righthand screw multiplication rules (2.1). With eliminating not important scaling factor $a$ we can write the following Clifford translation of any $G_3^+$ state operator $e^{I_{Plane}\varphi}$:

$$e^{-I_3\left(\frac{H}{|H|}\right)|H|t}e^{I_{Plane}\varphi}$$

which particularly satisfies (see [5], p.80) the Schrodinger equation for states as elements of $G_3^+$.

The first order linearization of the Hamiltonian exponent has the derivative by time $B_H e^{I_{Plane}\varphi}$ that obviously is the geometric algebra operator action on a state and corresponds to conventional quantum mechanics eigenstates/eigenvalues problem.

The above Clifford translation is product, composition, of states:

$$e^{-I_3\left(\frac{H}{|H|}\right)|H|t}e^{I_{Plane}\varphi} = e^{I_H|H|t}e^{I_{Plane}\varphi}, \qquad I_H = -I_3\left(\frac{H}{|H|}\right)$$

Since $I_H$ is not generally the same as (parallel to) $I_{Plane}$, the above composition of states has the following result (see, for example, [2]), with $\varphi$ and $I_{Plane}$ taken from (2.2) and (2.3), namely:

$$e^{I_H|H|t}e^{I_{Plane}\varphi} = \cos(|H|t)\cos\varphi + \sin(|H|t)\sin\varphi\,(I_H\cdot I_{Plane})$$
$$+ \sin(|H|t)\cos\varphi\,I_H + \cos(|H|t)\sin\varphi\,I_{Plane} + \sin(|H|t)\sin\varphi\,I_H\wedge I_{Plane} \quad (3.2)$$

For the general geometric algebra solution (1.2) of Maxwell equation, with the notations

$$\varphi^+ = \cos^{-1}\left(\frac{1}{\sqrt{2}}\cos\omega(t-[(I_3I_S)\cdot\vec{r}])\right), \qquad \varphi^- = \cos^{-1}\left(\frac{1}{\sqrt{2}}\cos\omega(t+[(I_3I_S)\cdot\vec{r}])\right)$$

$$I_{Plane}^+ = I_S \frac{\sin\omega(t-[(I_3I_S)\cdot\vec{r}])}{\sqrt{1+\sin^2\omega(t-[(I_3I_S)\cdot\vec{r}])}} + I_{B_0}\frac{\cos\omega(t-[(I_3I_S)\cdot\vec{r}])}{\sqrt{1+\sin^2\omega(t-[(I_3I_S)\cdot\vec{r}])}}$$
$$+ I_{E_0}\frac{\sin\omega(t-[(I_3I_S)\cdot\vec{r}])}{\sqrt{1+\sin^2\omega(t-[(I_3I_S)\cdot\vec{r}])}}$$

$$I_{Plane}^- = I_S \frac{\sin\omega(t+[(I_3I_S)\cdot\vec{r}])}{\sqrt{1+\sin^2\omega(t+[(I_3I_S)\cdot\vec{r}])}} + I_{B_0}\frac{\cos\omega(t+[(I_3I_S)\cdot\vec{r}])}{\sqrt{1+\sin^2\omega(t+[(I_3I_S)\cdot\vec{r}])}}$$
$$+ I_{E_0}\frac{\sin\omega(t+[(I_3I_S)\cdot\vec{r}])}{\sqrt{1+\sin^2\omega(t+[(I_3I_S)\cdot\vec{r}])}}$$

we get general state resulting from the Hamiltonian generated Clifford translation:

$$e^{I_H|H|t}\left(\lambda e^{I_{Plane}^+\varphi^+} + \mu e^{I_{Plane}^-\varphi^-}\right) = \cos(|H|t)\left(\lambda\cos\varphi^+ + \mu\cos\varphi^-\right)$$
$$+\sin(|H|t)\left(\lambda\sin\varphi^+(I_H\cdot I_{Plane}^+) + \mu\sin\varphi^-(I_H\cdot I_{Plane}^-)\right)$$
$$+\sin(|H|t)(\lambda\cos\varphi^+ + \mu\cos\varphi^-)I_H + \cos(|H|t)(\lambda\sin\varphi^+ I_{Plane}^+ + \mu\sin\varphi^- I_{Plane}^-)$$
$$+\sin(|H|t)\left(\lambda\sin\varphi^+(I_H\wedge I_{Plane}^+) + \mu\sin\varphi^-(I_H\wedge I_{Plane}^-)\right) \qquad (3.3)$$

As an example, let's find out what kind of Hamiltonian inclines bivector $I_S$ relative to wave vector $\hat{k}_+$. Take the plane of inclination orthogonal to $\vec{E}_0$, that's the plane of bivector $I_{E_0}$. The requirement is then that action of the Hamiltonian $e^{-I_3\left(\frac{H}{|H|}\right)|H|}$ on $I_S$ should be equal to rotation $e^{-I_{E_0}\gamma}I_S e^{I_{E_0}\gamma}$. Assume that the bivector basis used to lift the matrix form of Hamiltonian is taken as before $\{I_S, I_{B_0}, I_{E_0}\}$. Then the lift of (3.1), with $a=0$, is $I_3(bI_S + cI_{B_0} + dI_{E_0})$ and Clifford translation of $I_S$, $b^2+c^2+d^2\equiv\delta^2$, is $e^{\delta I_H}e^{\frac{\pi}{2}I_S}$, where $I_H = \frac{b}{\delta}I_S + \frac{c}{\delta}I_{B_0} + \frac{d}{\delta}I_{E_0}$. Using general formula (3.2) we get

$$e^{\delta I_H}e^{\frac{\pi}{2}I_S} = \sin\delta\,(I_H\cdot I_S) + \cos\delta\,I_S + \sin\delta\,I_H\wedge I_S = \frac{b}{\delta}\sin\delta + \cos\delta\,I_S + \frac{c}{\delta}\sin\delta\,I_{E_0} - \frac{d}{\delta}\sin\delta\,I_{B_0}$$

The result of the rotation $e^{-I_{E_0}\gamma}I_S e^{I_{E_0}\gamma}$ is: $e^{-I_{E_0}\gamma}I_S e^{I_{E_0}\gamma} = \cos 2\gamma\,I_S + \sin 2\gamma\,I_{B_0}$. Then it follows that:

$$b = c = 0, \qquad d = -\delta, \qquad \delta = 2\gamma,$$

and the geometric algebra Hamiltonian lift is $I_3(bI_S + cI_{B_0} + dI_{E_0}) = -2I_3\gamma I_{E_0}$. Its associated bivector is $I_H = I_{E_0}$ and the conventional quantum mechanics matrix form of this lift is:

$$\begin{pmatrix} 0 & 2i\gamma \\ -2i\gamma & 0 \end{pmatrix}$$

If we take inclination plane as the plane of $I_{B_0}$, instead of $I_{E_0}$, the result of rotation will be $\cos 2\gamma I_S - \sin 2\gamma I_{E_0}$, then

$$b = d = 0, \quad c = -\delta, \quad \delta = 2\gamma,$$

geometric algebra Hamiltonian lift is $I_3(bI_S + cI_{B_0} + dI_{E_0}) = -2I_3\gamma I_{B_0}$, and associated unit bivector is $I_H = I_{B_0}$ with the corresponding matrix Hamiltonian:

$$\begin{pmatrix} 0 & -2\gamma \\ -2\gamma & 0 \end{pmatrix}$$

In both cases of the Clifford translation of $I_S$, $e^{\delta I_H} e^{\frac{\pi}{2} I_S}$, geometrically rotation of $I_S$, we also get synchronic rotation by the same angle of the basis bivector $I_{B_0}$ in the plane of $I_{E_0}$ in the first case and rotation of the basis bivector $I_{E_0}$ in the plane of $I_{B_0}$ in the second case.

Let's calculate, as an example, the Berry potential in geometric algebra terms when the state (3.3) transforms due to Clifford translation by $e^{2\gamma I_{E_0}}$, rotation in the plane of $I_{E_0}$:

$$A(\gamma) = \left(\lambda e^{-I_{Plane}^+\varphi^+} + \mu e^{-I_{Plane}^-\varphi^-}\right) e^{-2\gamma I_{E_0}} I_{E_0} \frac{\partial}{\partial \gamma} e^{2\gamma I_{E_0}} \left(\lambda e^{I_{Plane}^+\varphi^+} + \mu e^{I_{Plane}^-\varphi^-}\right)$$

$$= \left(\lambda e^{-I_{Plane}^+\varphi^+} + \mu e^{-I_{Plane}^-\varphi^-}\right) e^{-2\gamma I_{E_0}} (-2 e^{2\gamma I_{E_0}}) \left(\lambda e^{I_{Plane}^+\varphi^+} + \mu e^{I_{Plane}^-\varphi^-}\right)$$

$$= (-2)\left(\lambda^2 + \mu^2 + \lambda\mu \left(e^{-I_{Plane}^+\varphi^+} e^{I_{Plane}^-\varphi^-} + e^{-I_{Plane}^-\varphi^-} e^{I_{Plane}^+\varphi^+}\right)\right)$$

Use general formula for geometrical product of states with different bivector planes [2]:

$$e^{I_{S_1}\varphi_1} e^{I_{S_2}\varphi_2} = \cos\varphi_1 \cos\varphi_2 + \sin\varphi_1 \cos\varphi_2 I_{S_1} + \cos\varphi_1 \sin\varphi_2 I_{S_2} + \sin\varphi_1 \sin\varphi_2 I_{S_1} I_{S_2}$$

$$= \cos\varphi_1 \cos\varphi_2 + \sin\varphi_1 \sin\varphi_2 (I_{S_1} \cdot I_{S_2}) +$$

$$\sin\varphi_1 \cos\varphi_2 I_{S_1} + \cos\varphi_1 \sin\varphi_2 I_{S_2} + \sin\varphi_1 \sin\varphi_2 I_{S_1} \wedge I_{S_2}$$

Then we have:

$$e^{-I_{Plane}^+\varphi^+} e^{I_{Plane}^-\varphi^-} + e^{-I_{Plane}^-\varphi^-} e^{I_{Plane}^+\varphi^+} = 2\cos\varphi^+ \cos\varphi^- - 2\sin\varphi^+ \sin\varphi^- (I_{Plane}^+ \cdot I_{Plane}^-)$$

Using from above

$$\cos\varphi^\pm = \frac{1}{\sqrt{2}} \cos\omega(t \mp [(I_3 I_S) \cdot \vec{r}]), \quad \sin\varphi^\pm = \frac{\sqrt{1+\sin^2\omega(t\mp[(I_3 I_S)\cdot\vec{r}])}}{\sqrt{2}},$$

along with formulas of $I_{Plane}^+$ and $I_{Plane}^-$, we get:

$$e^{-I_{Plane}^+\varphi^+} e^{I_{Plane}^-\varphi^-} + e^{-I_{Plane}^-\varphi^-} e^{I_{Plane}^+\varphi^+}$$
$$= \cos\omega(t - [(I_3 I_S)\cdot\vec{r}]) \cos\omega(t + [(I_3 I_S)\cdot\vec{r}])$$
$$- (-2\sin\omega(t - [(I_3 I_S)\cdot\vec{r}]) \sin\omega(t + [(I_3 I_S)\cdot\vec{r}])$$
$$- \cos\omega(t - [(I_3 I_S)\cdot\vec{r}]) \cos\omega(t + [(I_3 I_S)\cdot\vec{r}])) = 2\cos[2(I_3 I_S)\cdot\vec{r}]$$

So,

$$A(\gamma) = (-2)(\lambda^2 + \mu^2 + \lambda\mu\, 2\cos[2(I_3 I_S) \cdot \vec{r}])$$

If to assume $\lambda = \mu$ then, up to the field amplitude (equal values of $\vec{E}_0$ and $\vec{H}_0$), the geometric algebra Berry potential is:

$$A(\gamma) \sim - cos^2[(I_3 I_S) \cdot \vec{r}] \qquad (3.4)$$

In the case of Clifford translation by $e^{2\gamma I_{B_0}}$, rotation in the plane of $I_{B_0}$, the result remains the same.

The potential (3.4) does not depend of time and spreads through the whole three-dimensional space. The result is direct demonstration of instant nonlocality of states, operators, that can act on geometric algebra $G_3$ valued observables.

## 4. Conclusions

Two seminal ideas – variable and explicitly defined complex plane in three dimensions, and the $G_3^+$ states[3] as operators acting on observables – allow to put forth comprehensive and much more detailed formalism appropriate for quantum mechanics. The approach may be thought about as a far going geometric algebra generalization of some proposals for optical quantum computing. Maybe the most important thing is direct demonstration of instant state nonlocality.

Works Cited

[1] A. Soiguine, "Geometric Algebra, Qubits, Geometric Evolution, and All That," January 2015. [Online]. Available: http://arxiv.org/abs/1502.02169.

[2] A. Soiguine, "Polarizations as States and Their Evolution in Geometric Algebra Terms with Variable Complex Plane," *Journal of Applied Mathematics and Physics,* vol. 6, no. 4, 2018.

[3] A. Soiguine, "Quantum Computing with Geometric Algebra," in *Future Technologies Conference*, Vancouver, Canada, November 2017.

[4] A. M. Soiguine, "Complex Conjugation - Relative to What?," in *Clifford Algebras with Numeric and Symbolic Computations*, Boston, Birkhauser, 1996, pp. 284-294.

[5] A. Soiguine, Geometric Phase in Geometric Algebra Qubit Formalism, Saarbrucken: LAMBERT Academic Publishing, 2015.

---

[3] Good to remember that "state" and "wave function" are (at least should be) synonyms in conventional quantum mechanics